\documentclass[prl,reprint,amsmath,superscriptaddress]{revtex4-1}
\usepackage{hyperref}
\hypersetup{colorlinks,allcolors=blue,urlcolor=black,
			pdfauthor={Vincent Holten, J. Hrubý, M. E. H. van Dongen, and D. M. J. Smeulders},
            pdftitle={Comment on "Unraveling the 'Pressure Effect' in Nucleation"}}
\usepackage{mathptmx}
\usepackage[ansinew]{inputenc}
\usepackage{xspace}
\newcommand*{\ea}{\textit{et al.}\xspace}
\newcommand*{\peq}{p_{\text{eq}}}
\newcommand*{\pc}{p_{\text{c}}}
\newcommand*{\pt}{p_{\text{t}}}
\newcommand*{\liq}{_\ell}
\newcommand*{\vl}{v\liq}
\newcommand*{\vol}{v}
\newcommand*{\vap}{_{\text{v}}}
\newcommand*{\di}{\text{d}}
\newcommand*{\pure}{^0}
\newcommand*{\kB}{k_{\text{B}}}
\newcommand*{\kT}{\kB T}
\newcommand*{\DG}{\Delta G}
\newcommand*{\SW}{S_{\text{W}}}
\begin{document}
\title{Comment on ``Unraveling the `Pressure Effect' in Nucleation''}
\date{March 27, 2016}
\author{Vincent Holten}
\affiliation{Department of Mechanical Engineering, Eindhoven University of Technology, P.O. Box 513, 5600 MB Eindhoven, Netherlands}
\author{J. Hrubý}
\affiliation{Institute of Thermomechanics of the CAS, v. v. i., Dolejškova 5, CZ-182 00 Prague 8, Czech Republic}
\author{M. E. H. \surname{van Dongen}}
\affiliation{Department of Applied Physics, Eindhoven University of Technology, P.O. Box 513, 5600 MB Eindhoven, Netherlands}
\author{D. M. J. Smeulders}
\email[Corresponding author. Email: ]{D.M.J.Smeulders@tue.nl} \affiliation{Department of
Mechanical Engineering, Eindhoven University of Technology, P.O. Box 513, 5600 MB
Eindhoven, Netherlands} \maketitle In a 2008 Letter, Wedekind \ea \cite{wed08} discussed
the influence of an inert carrier gas on the vapor--liquid nucleation rate. They found an
additional ``pressure--volume work'' that is performed against the carrier gas, and also
quantified the nonisothermal effects arising from the carrier gas. We will argue that the
pressure--volume work term represents the influence of the carrier gas on phase
equilibrium itself. This term will not appear explicitly when a definition of the
supersaturation is used that is appropriate for high-pressure nucleation.

The presence of a background gas causes an increase in the equilibrium vapor pressure
(even when all substances are ideal), which is known as the Poynting effect
\cite{poynting1881}. Consider a vapor in equilibrium with its liquid phase in the
presence of a carrier gas. We will use the same notation as Wedekind \ea, that is, $p$ is
the vapor pressure, $\peq$ is the equilibrium vapor pressure, and $\pc$ is the carrier
gas pressure. In addition, we denote the total pressure as $\pt = p + \pc$. From an
integration of the Gibbs--Duhem equation $\di\mu = \vol\,\di p$ it follows that the
chemical potential of the liquid $\mu\liq$ at pressure $\pt$ is
\begin{equation}\label{eq:muliq}
\mu\liq (\pt) = \mu\liq\pure(\peq\pure) + \vl(\pt - \peq\pure),
\end{equation}
where superscript 0 denotes pure-component properties, and $\vl$ is the molecular volume.
Similarly, the chemical potential of the ideal vapor $\mu\vap$ at partial pressure $\peq$
and total pressure $\pt$ is
\begin{equation}\label{eq:muvap}
\mu\vap (\peq,\pt) = \mu\vap\pure(\peq\pure) + \kT\ln(\peq/\peq\pure),
\end{equation}
where $\kB$ is the Boltzmann constant and $T$ the temperature. Conditions of phase
equilibria for a pure vapor and for a vapor with carrier gas require
\begin{equation}
\mu\liq\pure(\peq\pure) = \mu\vap\pure(\peq\pure)\quad\text{and}\quad
\mu\liq (\pt) = \mu\vap (\peq,\pt),
\end{equation}
which yields for the equilibrium vapor pressure $\peq$ in the presence of a carrier gas and total pressure $\pt$
\begin{equation}
\peq = \peq\pure\exp\biggl[\frac{\vl(\pt-\peq\pure)}{\kT}\biggr],
\end{equation}
where the exponential is known as the Poynting factor.

The work of formation of a droplet of $n$ molecules is
\begin{equation}
\DG = n \bigl[\mu\liq(p\liq) - \mu\vap(p,\pt)\bigr] - n \vl (p\liq - \pt) + \gamma A,
\end{equation}
where $p\liq$ is the pressure in the droplet, $\gamma$ is the surface tension, and $A =
s_1 n^{2/3}$ is the area of the droplet with $s_1$ the surface area per monomer.
Analogously to Eqs.~(\ref{eq:muliq}) and (\ref{eq:muvap}) we obtain
\begin{align}
\mu\liq (p\liq) &= \mu\liq\pure(\peq\pure) + \vl(p\liq - \peq\pure)\\
\mu\vap (p,\pt) &= \mu\vap\pure(\peq\pure) + \kT\ln(p/\peq\pure),
\end{align}
and therefore
\begin{equation}
\DG = n \bigl[ \vl (\pt - \peq\pure) - \kT\ln(p/\peq\pure)\bigr]+ \gamma A,
\end{equation}
which corresponds to Eq. (5) of Wedekind \ea, with their definition of the
supersaturation, denoted here as $\SW = p/\peq\pure$. The term $n \vl (\pt - \peq\pure)$
includes the additional pressure--volume work against the carrier gas $W_{\text{c}} =
n\vl\pc$ that was introduced by Wedekind \ea

To incorporate carrier gas effects, the appropriate definition of supersaturation should be based on
the difference of the chemical potential of the vapor in the
actual state and at phase equilibrium at the actual total pressure as \cite{veh06p30,lui99a,pee01,fransen2015,heist1994}
\begin{equation}\label{eq:Sdef}
S = \exp\biggl[\frac{\mu\vap(p,\pt) - \mu\vap(\peq,\pt)}{\kT}\biggr].
\end{equation}
With this definition, the work of formation becomes
\begin{equation}\label{eq:work2}
\DG = - n \,\kT\ln S + \gamma A,
\end{equation}
and no pressure--volume term appears. Equation~(\ref{eq:work2}) was derived without
assuming ideal gas behavior and is therefore also valid for real gases and vapors. For
ideal gases, definition (\ref{eq:Sdef}) becomes
\begin{equation}\label{eq:S}
S = \frac{p}{\peq} =  \frac{p}{\peq\pure} \exp\biggl[-\frac{\vl(\pt - \peq\pure)}{\kT}\biggr].
\end{equation}
This definition of $S$ differs from $\SW$ by the Poynting factor. When comparing our
Eq.~(\ref{eq:work2}) with Eqs. (5) and (6) in Ref.~\citenum{wed08}, it can be seen that
the pressure--volume term occurs when the Poynting effect is not included in the
definition of the supersaturation. It should be noted that $S=1$ refers to phase
equilibrium at a given $\pc$ and $T$, while $\SW = 1$ does not. As a consequence, the
nucleation rate $J$ vanishes for $S \to 1$, but not for $\SW \to 1$. In fact, the
nucleation rate expression in Ref.~\citenum{wed08} $J_{pV}(\SW, \pc, T)$ requires a lower
limit of validity $\SW^{\text{min}}(\pc,T)$, while the lower bound for $J(S,T)$ equals
$S=1$ with our definition of the supersaturation.
\end{document}